\documentclass[useAMS,usenatbib]{mn2e}
\usepackage{amsmath,amssymb,color,graphicx}

\title{Baseline Dependent Averaging in Radio Interferometry}
\author[S.~J. Wijnholds, A.~G. Willis, S. Salvini]{S.~J. Wijnholds$^1$\thanks{E-mail: wijnholds@astron.nl}, A. G. Willis$^2$, S. Salvini$^3$\\
$^1$ASTRON, Oude Hoogeveensedijk 4, 7991 PD Dwingeloo, The Netherlands\\
$^2$National Research Council of Canada, P.O. Box 248, Penticton, BC V2A 6J9, Canada\\
$^3$OeRC, 7 Keble Road, Oxford, OX1 3QG, UK}

\def\bA{\mathbf{A}}
\def\bC{\mathbf{C}}
\def\bF{\mathbf{F}}
\def\bI{\mathbf{I}}
\def\bJ{\mathbf{J}}
\def\bM{\mathbf{M}}
\def\bR{\mathbf{R}}
\def\bW{\mathbf{W}}
\def\bX{\mathbf{X}}

\def\ba{\mathbf{a}}
\def\bl{\mathbf{l}}
\def\br{\mathbf{r}}
\def\bx{\mathbf{x}}

\def\bSigma{{\mbox{\boldmath{$\Sigma$}}}}

\def\btheta{{\mbox{\boldmath{$\theta$}}}}
\def\bxi{{\mbox{\boldmath{$\xi$}}}}
\def\bsigma{{\mbox{\boldmath{$\sigma$}}}}

\def\bone{\mathbf{1}}
\def\complex{\mathbb{C}}
\def\jcmplx{\mathrm{j}}
\def\real{\mathbb{R}}

\newcommand{\vectorize}[1]{\mathrm{vec} \left ( {#1} \right )}
\newcommand{\cov}[1]{\mathrm{cov} \left ( {#1} \right )}
\newcommand{\vecdiag}[1]{\mathrm{vecdiag} \left ( {#1} \right )}
\newcommand{\diag}[1]{\mathrm{diag} \left ( {#1} \right )}

\begin{document}

\pagerange{\pageref{firstpage}--\pageref{lastpage}} \pubyear{2016}

\maketitle

\label{firstpage}

\begin{abstract}
This paper presents a detailed analysis of the applicability and benefits of baseline dependent averaging (BDA) in modern radio interferometers and in particular the Square Kilometre Array (SKA). We demonstrate that BDA does not affect the information content of the data other than a well-defined decorrelation loss for which closed form expressions are readily available. We verify these theoretical findings using simulations. We therefore conclude that BDA can be used reliably in modern radio interferometry allowing a reduction of visibility data volume (and hence processing costs for handling visibility data) by more than 80\%.
\end{abstract}

\begin{keywords}
Instrumentation: interferometers, Methods: analytical, Methods: numerical, Techniques: interferometric
\end{keywords}

\section{Introduction}

Managing the large amount of visibility data is one of the biggest challenges in imaging science with modern radio interferometers, such as the Karl G. Jansky Very Large Array (JVLA, \cite{Perley2009ProcIEEE}), the Murchison Widefield Array (MWA, \cite{Tingay2013PASA}), the Low Frequency Array (LOFAR, \cite{Haarlem2013AandA}) and in particular the future Square Kilometre Array (SKA, \cite{Dewdney2009ProcIEEE}). The visibility data volumes are driven upwards by the small channel bandwidth and short correlator dump times required to limit smearing and decorrelation effects on the longest baselines and at the edges of the field-of-view to an acceptable level \citep{Bridle1999}. Since this small channel width and high time resolution are not required on the shorter baselines, \cite{Cotton2009Obit} and \cite{Skipper2014} have considered baseline dependent averaging (BDA) in the context of the VLA and the SKA respectively. BDA has also been used for MWA, which initially had to operate with stringent storage capacity restrictions \citep{Mitchell2008JSTSP}. Besides data compression, BDA can also be used to shape the field-of-interest \citep{Atemkeng2016MNRAS}.

Time and frequency averaging causes smearing effects that do not have a convolutional nature in the image plane as explained by \cite{Bridle1999}. \cite{Cotton2009Obit} shows that this has detrimental effects on the achievable dynamic range, which is of particular concern in the context of modern radio interferometers. He alludes to the possibility to make first order corrections for these smearing effects at the expense of an increase in computing requirements. Since LOFAR experience shows that imaging pipelines for SKA may actually be limited by data handling instead of computing, this increase in required computing may be acceptable if it makes the visibility data volumes more manageable. The question then still remains, whether BDA poses a fundamental limit on dynamic range if the appropriate corrections can be made during the imaging process. This is the first question considered in this paper. It will be addressed by an analysis of information content of visibilities with and without application of BDA. This analysis is presented in Sec.~\ref{sec:theory}.

The length of the solution interval for calibration is another concern in the context of BDA, in particular for time averaging. In this case, the integration time on the shortest baselines may become longer than the length of the solution interval required to keep track of the fastest changes in the observing system. The obvious solution would be to limit the maximum allowed integration time on any baseline to an interval that is shorter than the solution interval used for calibration. The problem with this approach is, that the length of the solution interval may be chosen differently during post-processing of the data to deal with, for example, large temperature variations or worse ionospheric or tropospheric conditions than expected. This risk can be mitigated by the Compress-Expand-Compress (CEC) method described by \cite{Salvini2017URSI}. For convenience of the reader, we recapitulate the main results presented in \cite{Salvini2017URSI} in Sec.~\ref{sec:CEC}. Based on these results, we will assume that we are justified in averaging the data to the extend that we do in our analysis and simulations.

We validate our analysis using simulations in which we study the difference between images obtained from full-resolution visibilities and visibilities obtained after applying BDA. These simulations, which are presented in Sec.~\ref{sec:sims}, are set up in such a way that the impact of BDA is isolated from other effects, such as incomplete sky models or imperfect instrument models, that we usually face when making high dynamic range images. These simulations confirm the main conclusion from our theoretical analysis that BDA is not likely to have a detrimental effect on imaging performance besides a well-understood decorrelation effect.

%Our simulations indicate that the CEC method can successfully recover short-term gain variations from data that would normally be considered to have been integrated too long on the shortest baselines.

\section{Theoretical analysis}
\label{sec:theory}

\subsection{Mathematical description of BDA}
\label{ssec:BDA_description}

A correlator usually correlates narrowband time-series data from $P$ receiving elements, which can either be aperture array stations or dishes. We can write the signal from the $p$th receiving element at frequency $\nu$ as $x_p \left ( \left ( l N + n \right ) T, \nu \right )$, where $T$ is the sample interval, $n$ is the sample index in the $l$th short term integration (STI) and $N$ is the length of a short term integration, which is determined by the correlator dump time needed to accurately sample the visibilities on the longest baselines. For Nyquist sampled input signals, $N = B \tau$, where $B$ is the channel bandwidth and $\tau$ is the integration time of a STI. The corresponding time samples for all $P$ receiving elements can be stacked into the array signal vector
\begin{equation}
\bx \left ( \left ( l N + n \right ) T, \nu \right ) = \left [
\begin{array}{c}
x_1 \left ( \left ( l N + n \right ) T, \nu \right )\\
\vdots\\
x_P \left ( \left ( l N + n \right ) T, \nu \right )
\end{array} \right ] \in \complex^{P \times 1}.
\end{equation}

The time samples for the $l$th STI can be collected in the matrix
\begin{equation}
\bX_{l,\nu} = \left [ \bx_\nu \left ( \left ( l N + 1 \right ) T \right ), \cdots, \bx_\nu \left ( \left ( l N + N \right ) T \right ) \right ] \in \complex^{P \times N},
\end{equation}
where we have changed the frequency $\nu$ into an indexing parameter, since BDA can not only be applied along the time axis, but also along the frequency axis. Using these samples, a correlator that does not use BDA, will estimate the visibility matrix or array covariance matrix
\begin{equation}
\widehat{\bR}_{l,\nu} = \frac{1}{N} \bX_{l, \nu} \bX_{l,\nu}^H \in \complex^{P \times P}
\end{equation}
for the $l$th STI. Note that the autocorrelations are included in this matrix. Although these are often discarded in radio astronomical data reduction, they are essential for our statistical analysis, because the autocorrelations measure the system temperature.

Each of these visibility matrices can be vectorized by stacking its columns to form visibility vectors $\widehat{\br}_{l,\nu} = \vectorize{\widehat{\bR}_{l,\nu}} \in \complex^{P^2 \times 1}$. We can replace the indexing parameters $l$ and $\nu$ to index STIs over time and frequency respectively by a single index parameter $k$. Let us assume that $K$ STIs, which can be collected over time or frequency or both, are combined in a self-calibration and imaging process to estimate calibration and image parameters stacked in the parameter vector $\btheta$. After defining the raw visibility data vector as
\begin{equation}
\widehat{\br} = \left [ \widehat{\br}_1^T, \cdots, \widehat{\br}_K^T \right ] \in \complex^{KP^2 \times 1},
\end{equation}
the self-calibration problem can be formulated as
\begin{equation}
\widehat{\btheta} = \underset{\theta}{\mathrm{argmin}} \left \| \widehat{\br} - \br \left ( \btheta \right ) \right \|^2,
\end{equation}
where $\br \left ( \btheta \right )$ represents the parameterized measurement equation.

The length of the raw visibility data vector can be reduced from $K P^2$ to $M \leq K P^2$ by averaging the raw visibilities on some baselines. Integration over multiple STIs by \emph{summation} can be described by
\begin{equation}
\br_\mathrm{sum} = \bI_\mathrm{s} \br \in \complex^{M \times 1}.
\end{equation}
The selection matrix $\bI_\mathrm{s} \in \real^{M \times K P^2}$ is a sparse matrix with only a single element equal to unity in each column. Each row contains at least a single element equal to unity, but may contain multiple elements with value 1 depending on the number of raw visibility samples being added into a single summed visibility. This selection matrix can be partitioned in selection matrices for each individual STI $\bI_{\mathrm{s},k} \in \real^{M \times P^2}$ as
\begin{equation}
\bI_\mathrm{s} = \left [ \bI_{\mathrm{s},1}, \cdots, \bI_{\mathrm{s},K} \right ]. \label{eq:partitioning_Is}
\end{equation}
The selection matrix for a single STI may have one or more rows containing only zeros signifying that that particular STI does not contribute to the summed visibility associated with that row.

The \emph{averaged} visibilities $\br_\mathrm{ave}$ follow from the summed visibilities by dividing each element of $\br_\mathrm{sum}$ by the number of raw visibilities that contributed to that element. This can be done using the diagonal weighting matrix
\begin{equation}
\bW = \left ( \bI_\mathrm{s} \bI_\mathrm{s}^T \right )^{-1} \in \real^{M \times M} \label{eq:W}
\end{equation}
to obtain
\begin{equation}
\br_\mathrm{ave} = \bW \br_\mathrm{sum} = \bW \bI_\mathrm{s} \br \in \complex^{M \times 1}. \label{eq:rave_from_r}
\end{equation}

Ignoring decorrelation effects, which will be discussed in Sec.~\ref{ssec:decorrelation}, the visibility values that are averaged together, are assumed to measure the same physical quantity, i.e., the expected value of these data values is assumed to be the same. This assumption means that the expected value of $\br$ can be obtained from the expected value of $\widehat{\br}_\mathrm{ave}$ by
\begin{equation}
\br = \mathcal{E} \left \{ \widehat{\br} \right \} = \bI_\mathrm{s}^H \br_\mathrm{ave}.
\label{eq:rave_to_r}
\end{equation}
Strictly speaking, this equation is not valid for realized values, because the noise on the non-averaged visibilities is averaged in the averaged visibilities and can therefore not be recovered. However, as long as we work with expected values, we can use this equation to transform from the space of averaged visibilities to the space of non-averaged visibilities. Interestingly, \eqref{eq:rave_to_r} describes the most basic expand step in the Compress-Expand-Compress approach proposed by \cite{Salvini2017URSI} and summarized in Sec.~\ref{sec:CEC}.

\subsection{Information content of averaged visibilities}

The Cramer-Rao bound (CRB) provides a lower bound on the covariance of all estimated parameters for an unbiased estimator \citep{Kay1993}. The CRB is the inverse of the Fisher Information Matrix (FIM), which measures the amount of information a given data vector contains about an unknown parameter vector. In this section, we assess the preservation of information when applying BDA by comparing the CRB for estimation of $\br_\mathrm{ave}$ from $\br$ with the covariance of $\br_\mathrm{ave}$ obtained by propagating the noise on the raw visibilities $\br$ to the averaged visibilities.

The FIM is defined as \citep{Kay1993}
\begin{equation}
\bF = \bJ^H \cov{\br}^{-1} \bJ,
\end{equation}
where $\bJ = \partial \br / \partial \btheta^T$ is a Jacobian matrix containing the partial derivatives for each element of $\br$ to all parameters in the parameter vector $\btheta$ and $\cov{\br}$ is the covariance matrix of the raw visibility vector $\br$. For our analysis, we have $\btheta = \br_\mathrm{ave}$, so, using \eqref{eq:rave_to_r}, the Jacobian matrix becomes
\begin{equation}
\bJ = \frac{\partial \br}{\partial \br_\mathrm{ave}^T} = \bI_\mathrm{s}^H.
\end{equation}
The CRB for estimating $\br_\mathrm{ave}$ from $\br$ thus becomes
\begin{equation}
\bC = \bF^{-1} = \left ( \bI_\mathrm{s} \cov{\br}^{-1} \bI_\mathrm{s}^H \right )^{-1}. \label{eq:CRB_rave_from_r}
\end{equation}

If the STIs are uncorrelated, i.e., if the time and frequency samples can be considered as independent measurements, the covariance of $\br$ can be written as
\begin{equation}
\cov{\br} = \left [ \begin{array}{ccc}
\cov{\br_1} & &\\
& \ddots &\\
& & \cov{\br_K} \end{array} \right ],
\end{equation}
i.e., as a block diagonal matrix with the covariance matrices of the individual STIs on the main diagonal. Using the fact that the inverse of a block diagonal matrix is a block diagonal matrix formed by the inverse of the individual blocks and the fact that $\bI_\mathrm{s}$ can be partitioned per STI as shown in \eqref{eq:partitioning_Is}, we can write \eqref{eq:CRB_rave_from_r} as
\begin{equation}
\bC = \left ( \sum_{k=1}^K \bI_{\mathrm{s},k} \cov{\br_k}^{-1} \bI_{\mathrm{s},k}^H \right )^{-1}. \label{eq:CRB_rave_from_r_STI}
\end{equation}

Using standard error propagation formulas, we find that the covariance of $\br_\mathrm{ave}$ obtained using the BDA procedure described earlier is given by
\begin{eqnarray}
\cov{\br_\mathrm{ave}} & = & \left ( \frac{\partial \br_\mathrm{ave}}{\partial \br^T} \right ) \cov{\br} \left ( \frac{\partial \br_\mathrm{ave}}{\partial \br^T} \right )^H \nonumber\\
& = & \bW \bI_\mathrm{s} \cov{\br} \bI_\mathrm{s}^H \bW^H \nonumber\\
& = & \bW \left ( \sum_{k=1}^K \bI_{\mathrm{s},k} \cov{\br_k} \bI_{\mathrm{s},k}^H \right ) \bW^H. \label{eq:cov_rave_from_r}
\end{eqnarray}

Comparison of the CRB for estimating $\br_\mathrm{ave}$ from $\br$ given by \eqref{eq:CRB_rave_from_r_STI} with the realised covariance of $\br_\mathrm{ave}$ given by \eqref{eq:cov_rave_from_r} shows that they are not the same in general. This may come as a surprise, since we are used to increasing our integration time to reduce the noise on our data. It is instructive to demonstrate that integration over the same number of STIs on all baselines is one of the possible ways to make the covariance of $\br_\mathrm{ave}$ equal to the CRB, thereby confirming our intuition that averaging of data to reduce the noise per data value does not affect our ability to estimate parameters from those data. If we integrate over $K$ STIs on all baselines, we get $\bI_{\mathrm{s},k} = \bI \in \real^{P^2 \times P^2}$ and $\bW = \frac{1}{K} \bI \in \real^{P^2 \times P^2}$, where $\bI$ denotes the identity matrix of appropriate size. This simplifies the CRB to
\begin{equation}
\bC = \frac{1}{K} \cov{\br_1},
\end{equation}
where we assumed that the covariance is the same for all STIs, and \eqref{eq:cov_rave_from_r} to
\begin{equation}
\cov{\br_\mathrm{ave}} = \frac{1}{K} \cov{\br_1},
\end{equation}
which is equal to the CRB as expected. The factor $1/K$ confirms our common knowledge that the noise reduces with the square root of the number of samples when we integrate (remember that the CRB represents the (co)variance of the estimated parameters while noise is usually expressed in standard deviations).

Another special case, in which the covariance of $\br_\mathrm{ave}$ becomes (approximately) identical to the CRB, is the regime in which the instantaneous SNR is low. In this regime, the system temperature of the receiving elements (aperture array stations or dishes) is significantly higher than the source temperature. Therefore, the power measured in the autocorrelations is significantly higher than the power observed in the crosscorrelations, so we may assume that $\bR_k \approx \sigma_n \bI$ for an array of identical receiving elements. If all received source and noise signals can be described as independent and identically distributed Gaussian noise, $\cov{\br_k} = \frac{1}{N} \sigma^2 \bI \in \complex^{P^2 \times P^2}$. This gives $\cov{\br} = \frac{1}{N} \sigma^2 \bI \in \complex^{KP^2 \times KP^2}$. Substituting this in \eqref{eq:CRB_rave_from_r} gives
\begin{equation}
\bC = \left ( \bI_\mathrm{s} \left ( \frac{\sigma^2}{N} \bI \right )^{-1} \bI_\mathrm{s}^H \right )^{-1} = \left ( \frac{N}{\sigma^2} \bI_\mathrm{s} \bI_\mathrm{s}^H \right )^{-1} = \frac{\sigma^2}{N} \bW, \label{eq:CRB_low_SNR}
\end{equation}
where we used \eqref{eq:W} in the last step.

Using the same result in \eqref{eq:cov_rave_from_r}, we find
\begin{equation}
\cov{\br_\mathrm{ave}} = \bW \bI_\mathrm{s} \frac{\sigma^2}{N} \bI \bI_\mathrm{s}^H \bW = \frac{\sigma^2}{N} \bW \bI_\mathrm{s} \bI_\mathrm{s}^H \bW = \frac{\sigma^2}{N} \bW,
\end{equation}
which is equal to the CRB derived in \eqref{eq:CRB_low_SNR}. We can therefore conclude that the BDA scheme described in Sec.~\ref{ssec:BDA_description} works well in cases where the instantaneous SNR, i.e., the SNR before integration, of the observed sources is very low. However, in cases in which the instantaneous SNR is close to unity or even higher, a more advanced BDA scheme is required. Unfortunately, such a BDA scheme will require significantly more computing resources. Since the sources in the high-SNR case are observed with very high SNR, it is unlikely that a small loss of information will lead to a significant reduction in the scientific output of such observations. This intuition is corroborated by the analysis presented in Sec.~\ref{ssec:imaging} that demonstrates that this information loss does not affect our imaging capability directly.

The conclusion that some information is lost when applying BDA as described in Sec.~\ref{ssec:BDA_description} in the high-SNR regime obviously raises the following questions:
\begin{enumerate}
\item How much information is lost? In other words, how much does the noise on the averaged visibilities increase compared to the optimal case in which no information is lost?
\item Is there a BDA scheme that does not incur an information loss and, if so, what is this scheme?
\end{enumerate}
These issues are addressed in the next two subsections. These subsections aim to provide some more insight into the information loss issue identified above, but do not affect the conclusions already drawn or provide details that are used later in the paper. The reader may therefore skip these two subsections if desired.

\subsubsection{A statistically optimal BDA scheme}

In this section, we derive a statistically efficient estimator to estimate $\br_\mathrm{ave}$ from $\br$. A statistically efficient estimator is unbiased and provides the lowest possible covariance on the estimated parameters. A maximum-likelihood (ML) estimator is such an estimator. It is known that optimization of the weighted least squares cost function with covariance matched weighting leads to estimates that are asymptotically, i.e., for a large number of samples, equal to ML estimates \citep{Ottersten1998DSP}. We thus want to solve
\begin{equation}
\widehat{\br}_\mathrm{ave} = \underset{\br_\mathrm{ave}}{\mathrm{argmin}} \left \| \cov{\br}^{-1/2} \left ( \widehat{\br} - \bI_\mathrm{s}^H \br_\mathrm{ave} \right ) \right \|^2.
\end{equation}

The solution to this problem is well known and is given by
\begin{equation}
\widehat{\br}_\mathrm{ave} = \left ( \cov{\br}^{-1/2} \bI_\mathrm{s} \right )^\dagger \cov{\br}^{-1/2} \widehat{\br},
\end{equation}
where $^\dagger$ denotes the pseudo-inverse. Using the Moore-Penrose inverse, we can write this as
\begin{equation}
\widehat{\br}_\mathrm{ave} = \left ( \bI_\mathrm{s} \cov{\br}^{-1} \bI_\mathrm{s}^H \right )^{-1} \bI_\mathrm{s} \cov{\br}^{-1} \widehat{\br}. \label{eq:rave_from_r_opt}
\end{equation}

Using the standard error propagation formula, as we did before to obtain \eqref{eq:cov_rave_from_r}, we find,
\begin{eqnarray}
\cov{\br_\mathrm{ave}} & = & \left ( \bI_\mathrm{s} \cov{\br}^{-1} \bI_\mathrm{s}^H \right )^{-1} \bI_\mathrm{s} \cov{\br}^{-1} \cov{\br} \times \nonumber\\
& & \cov{\br}^{-1} \bI_\mathrm{s}^H \left ( \bI_\mathrm{s} \cov{\br}^{-1} \bI_\mathrm{s}^H \right )^{-1} \nonumber\\
& = & \left ( \bI_\mathrm{s} \cov{\br}^{-1} \bI_\mathrm{s}^H \right )^{-1}, \label{eq:cov_rave_from_r_opt}
\end{eqnarray}
which is equal to the CRB for estimating $\br_\mathrm{ave}$ from $\br$ derived in \eqref{eq:CRB_rave_from_r}. This shows that it is possible to develop a BDA scheme that incurs no information loss (besides the expected decorrelation discussed later). However, the required weighting is data dependent, which means that the weighting matrix cannot be precomputed, and is computationally expensive, even if the block-diagonal structure of $\cov{\br}$ is exploited. Since information loss mainly occurs in high-SNR scenarios, it is probably acceptable and applying such an advanced BDA scheme is probably not worth the investment. This conclusion is examined in more detail below.

\subsubsection{Quantification of information loss}

To assess the impact of the instantaneous SNR on estimation performance, we conduct a simulation in which we estimated $\widehat{\br}_\mathrm{ave}$ from $\widehat{\br}$ using the optimal estimator described by \eqref{eq:rave_from_r_opt} as well as using the proposed BDA scheme described by \eqref{eq:rave_from_r}.

For this simulation, we use a simple visibility model consisting of a single source with power $\sigma_\mathrm{s} = 1$ in the phase center and uncorrelated additive noise with power $\sigma_\mathrm{n} = \sigma_\mathrm{s} / \mathrm{SNR}$ per receiving element. For simplicity, we assume that the source model is constant over all STIs. We thus have
\begin{equation}
\bR_k = \sigma_\mathrm{s} \bone \bone^H + \sigma_\mathrm{n} \bI \in \complex^{P \times P} ~~\forall k,
\end{equation}
where $\bone$ denotes a column vector filled with ones of appropriate length. Note that this model is independent of the array layout as the source is in the phase center. Assuming that the signals, from which $\bR_k$ is estimated, can all be described by independent and identically distributed complex Gaussian signals, we have (see, for example, appendix C in \cite{Boonstra2005PhD})
\begin{equation}
\cov{\br_k} = \frac{1}{N} \overline{\bR}_k \otimes \bR_k,
\end{equation}
where overbar denotes conjugation and $\otimes$ denotes the Kronecker product of two matrices.

For our comparison, we calculated
\begin{enumerate}
\item the CRB for estimating $\br_\mathrm{ave}$ from $\br$ using \eqref{eq:CRB_rave_from_r};
\item the covariance of $\br_\mathrm{ave}$ for the BDA scheme described by \eqref{eq:rave_from_r} using the standard error propagation formula as in \eqref{eq:cov_rave_from_r};
\item the covariance of $\br_\mathrm{ave}$ when $\br_\mathrm{ave}$ is estimated using the optimal BDA scheme described in \eqref{eq:rave_from_r_opt}.
\end{enumerate}

In our simulations, we varied the SNR from $10^{-3}$ to 1 in 19 steps (20 SNR values in total) equidistantly spaced on a logarithmic scale. For clarity, we used a simple 10-element uniform linear array for this example instead of the SKA-mid configuration that will be used later in this paper. The maximum integration time was set to 16 STIs and the number of values averaged was doubled every time the baseline length was halved.

\begin{figure}
\centering
\includegraphics[width=0.9\columnwidth]{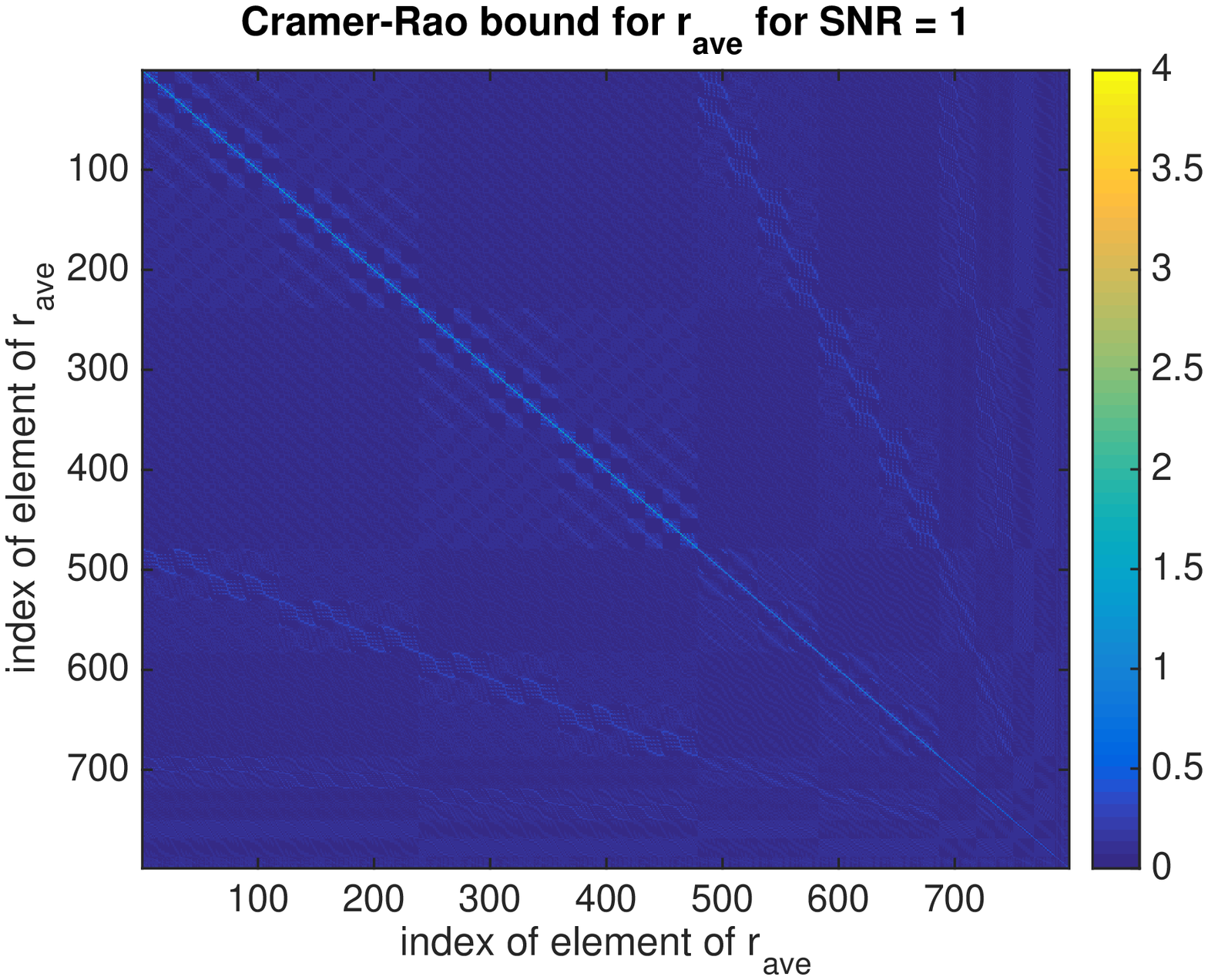}
\includegraphics[width=0.9\columnwidth]{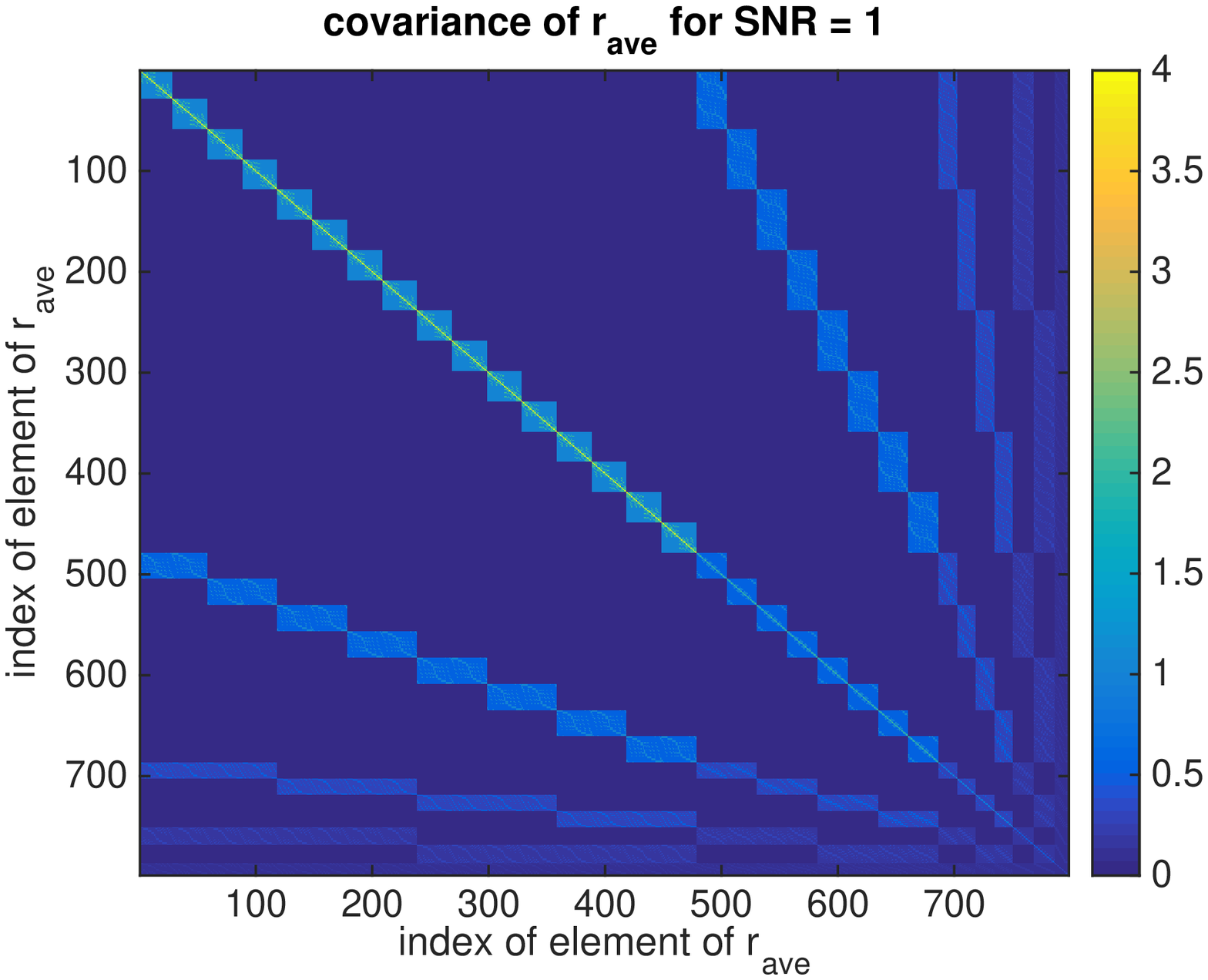}
\caption{CRB (top) and covariance of $\br_\mathrm{ave}$ when estimated using the normal BDA scheme described by \eqref{eq:rave_from_r} for an SNR of unity. \label{fig:cov_rave}}
\end{figure}

Figure \ref{fig:cov_rave} compares the covariance of $\br_\mathrm{ave}$ estimated using the non-optimal BDA scheme described by \eqref{eq:rave_from_r} with the CRB. The covariance of $\br_\mathrm{ave}$ using the optimal weighting described by \eqref{eq:rave_from_r_opt} was equal to the CRB as expected and is therefore not shown separately. The covariance of $\br_\mathrm{ave}$ estimated using \eqref{eq:rave_from_r} is not only significantly higher than the CRB, indicating a lower estimation accuracy, but also has a different structure. This suggests a significant loss and change of information content in the data.

\begin{figure}
\centering
\includegraphics[width=0.9\columnwidth]{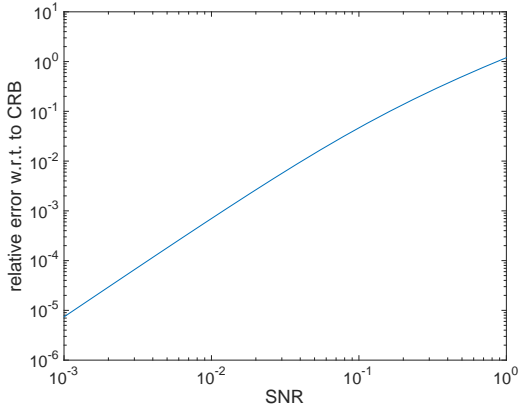}
\caption{Mean relative increase of the variance relative to the CRB for $\br_\mathrm{ave}$ obtained after BDA as function of SNR as calculated using Eq.~\eqref{eq:mean_relative_error}. \label{fig:cov_rave_relative}}
\end{figure}

To assess the impact of SNR, we computed the mean relative increase of the variance relative to the CRB for $\br_\mathrm{ave}$:
\begin{equation}
\epsilon_\mathrm{mean} = \frac{1}{M} \sum_{m=1}^M \frac{\left | \left [ \vecdiag{\bC} \right ]_m - \left [ \vecdiag{\cov{\br_\mathrm{ave}}} \right ]_m \right |}{\left [ \vecdiag{\bC} \right ]_m}, \label{eq:mean_relative_error}
\end{equation}
where $\left [ \cdot \right ]_m$ denotes the $m$th element of the vector. Figure \ref{fig:cov_rave_relative} shows this measure as function of SNR. The results confirm our theoretical finding that the difference between the covariance of $\br_\mathrm{ave}$ obtained using BDA and the CRB only has a significant effect in high-SNR regimes. Since the sources are then observed with a high SNR, a small increase in the measurement error on the data values may be perfectly acceptable.

To get an intuition for the typical instantaneous SNR in a radio astronomical observation, let us consider an observation with LOFAR on  Hercules A, one of the brightest astronomical sources in the northern hemisphere. Hercules A has a flux of 325 Jy at 178 MHz \citep{Bennett1962}, while the source equivalent flux density (SEFD) of a single LOFAR HBA core station is about 3.5 kJy \citep{Haarlem2013AandA}. Based on Fig.~\ref{fig:cov_rave_relative}, we may expect a deviation of order 1\% from the CRB, which is effectively negligible given the typical contributions from other observing and data reduction errors.

\subsection{Impact on imaging}
\label{ssec:imaging}

In our analysis in the previous section, we found that the BDA scheme defined by \eqref{eq:rave_from_r} incurs a (small) loss of information compared to optimal estimation of $\br_\mathrm{ave}$ from $\br$. Unfortunately, the CRB (or FIM) of the averaged visibilities does not tell us whether the information content of the raw visibilities is preserved in optimal averaging. In this section, we address this issue by considering the ability to reconstruct an image from the raw (unaveraged) visibilities and from averaged visibilities.

The sky can be discretized in $Q$ directions. If a plane wave with power $\sigma_q$ is impinging on the array from the $q$th direction, the visibility response of a perfectly calibrated array is given by \citep{Veen2013}
\begin{equation}
\bR_{k,q} = \sigma_q \ba_{k,q} \ba_{k,q}^H, \label{eq:Rkq}
\end{equation}
where $\ba_{k,q} = \left [ e^{2\pi\jcmplx \bxi_{k,1} \cdot \bl_q/\lambda}, \cdots, e^{2\pi\jcmplx \bxi_{k,P} \cdot \bl_q / \lambda} \right ]^T \in \complex^{P \times 1}$ is the array response vector. The vector $\bl_q$ denotes the unit vector pointing in the $q$th direction and $\bxi_{k,p}$ is the position vector of the $p$th receiving element in the array during the $k$th snapshot. Vectorization of \eqref{eq:Rkq} and forming a superposition of all source and noise signals gives
\begin{equation}
\br_k = \left ( \sum_{q=1}^Q \left ( \overline{\ba}_{k,q} \otimes \ba_{k,q} \right ) \sigma_q \right ) + \vectorize{\bSigma_\mathrm{n}}, \label{eq:rkq}
\end{equation}
where $\bSigma_\mathrm{n}$ denotes the noise covariance matrix. If the receiver noise powers are uncorrelated, the noise covariance matrix is diagonal, i.e., $\bSigma_\mathrm{n} = \diag{\bsigma_\mathrm{n}}$. Stacking the array response vectors $\ba_{k,q}$ in an array response matrix $\bA_k = \left [ \ba_{k,1}, \cdots, \ba_{k,Q} \right ] \in \complex^{P \times Q}$ and the source powers $\sigma_q$ in a vector $\bsigma_\mathrm{s} = \left [ \sigma_1, \cdots \sigma_Q \right ]^T$, we can write \eqref{eq:rkq} as
\begin{eqnarray}
\br_k & = & \left ( \overline{\bA}_k \circ \bA_k \right ) \bsigma_\mathrm{s} + \left ( \bI \circ \bI \right ) \bsigma_\mathrm{n} \nonumber\\
& = & \left [ \begin{array}{cc} \overline{\bA}_k \circ \bA_k & \bI \circ \bI \end{array} \right ] \left [ \begin{array}{c} \bsigma_\mathrm{s}\\ \bsigma_\mathrm{n} \end{array} \right ] \nonumber\\
& = & \bM_k \btheta,
\end{eqnarray}
where $\circ$ denotes the Khatri-Rao or column-wise Kronecker product of two matrices. We have also introduced the parameter vector $\btheta = \left [ \bsigma_\mathrm{s}^T, \bsigma_\mathrm{n}^T \right ]^T$ and the measurement matrix $\bM_k \in \complex^{P^2 \times \left ( Q + P \right )}$. Stacking all snapshots in a single visibility vector, we get
\begin{equation}
\br = \left [ \begin{array}{c} \bM_1\\ \vdots\\ \bM_K \end{array} \right ] \btheta = \bM \btheta.
\end{equation}

In our image reconstruction process, we want to solve
\begin{equation}
\widehat{\btheta} = \underset{\btheta}{\mathrm{argmin}} \left \| \cov{\br}^{-1/2} \left ( \widehat{\br} - \bM \btheta \right ) \right \|^2,
\end{equation}
where we have assumed covariance matched weighting to obtain an estimate that is asymptotically equivalent to the ML estimate \citep{Ottersten1998DSP}. The standard solution to this problem is given by
\begin{eqnarray}
\widehat{\btheta} & = & \left ( \bM^H \cov{\br}^{-1} \bM \right )^{-1} \bM^H \cov{\br}^{-1} \widehat{\br} \nonumber\\
& = & \bM_\mathrm{im} \widehat{\br}, \label{eq:image_reconstruction}
\end{eqnarray}
i.e., the image reconstruction process can be described by a matrix $\bM_\mathrm{im} \in \complex^{\left ( Q + P \right ) \times KP^2}$ that maps the measured visibilities on the parameters describing the image and the instrumental noise. It is interesting to note that this simple equation represents the imaging process including deconvolution in the case of weighted least squares image reconstruction while it can also represent a simple DFT imager or creation of a dirty image by gridding followed by a fast Fourier transform \citep{Veen2013}. In other words, \eqref{eq:image_reconstruction} describes a whole class of image reconstruction methods in which the visibilities can be described as a linear superposition of sources signals.

Using standard error propagation, as before, the covariance of $\btheta$ is given by
\begin{eqnarray}
\cov{\btheta} & = & \left ( \frac{\partial \btheta}{\partial \br^T} \right ) \cov{\br} \left ( \frac{\partial \btheta}{\partial \br^T} \right )^H \nonumber\\
& = & \bM_\mathrm{im} \cov{\br} \bM_\mathrm{im}^H.
\end{eqnarray}

If we use BDA using a scheme commensurate with \eqref{eq:rave_from_r} and apply the image reconstruction process described by $\bM_\mathrm{im}$ to the full resolution visibility vector restored from $\br_\mathrm{ave}$ using \eqref{eq:rave_to_r}, we obtain the reconstructed image give by
\begin{equation}
\widehat{\btheta}_\mathrm{ave} = \bM_\mathrm{im} \bI_\mathrm{s}^H \widehat{\br}_\mathrm{ave}.
\end{equation}
In this case, the covariance of the reconstructed parameters will be
\begin{equation}
\cov{\btheta_\mathrm{ave}} = \bM_\mathrm{im} \bI_\mathrm{s}^H \cov{\br_\mathrm{ave}} \bI_\mathrm{s} \bM_\mathrm{im}^H.
\end{equation}
Substitution of \eqref{eq:cov_rave_from_r} gives
\begin{equation}
\cov{\btheta_\mathrm{ave}} = \bM_\mathrm{im} \bI_\mathrm{s}^H \bW \bI_\mathrm{s} \cov{\br} \bI_\mathrm{s}^H \bW \bI_\mathrm{s} \bM_\mathrm{im}^H.
\end{equation}

In this derivation, we reconstruct the non-averaged visibilities using \eqref{eq:rave_to_r}. This operation simply assigns each value in $\br_\mathrm{ave}$ to the entries of the reconstructed non-averaged visibility vector $\widetilde{\br}$ corresponding to the entries of $\br$ that were averaged to obtain that value in $\br_\mathrm{ave}$. If $\br$ describes a visibility \emph{model} commensurate with BDA, i.e., if $\br$ contains no noise and entries that are averaged in BDA have the same value, we will have $\widetilde{\br} = \br$. Since $\br_\mathrm{ave} = \bW \bI_\mathrm{s} \br$, this implies that
\begin{equation}
\widetilde{\br} = \br = \bI_\mathrm{s}^H \bW \bI_\mathrm{s} \br.
\end{equation}
This shows that multiplication with $\bI_\mathrm{s}^H \bW \bI_\mathrm{s}$ has the same effect as multiplication with the identity matrix for nosie-free vectors, such as vectors containing expected values or model values, that have a structure commensurate with BDA. If each column of $\bM_\mathrm{im}^H$ has that property, it thus follows that
\begin{equation}
\bI_\mathrm{s}^H \bW \bI_\mathrm{s} \bM_\mathrm{im}^H = \bM_\mathrm{im}^H.
\end{equation}
In this case, it is easy to show that
\begin{equation}
\cov{\btheta_\mathrm{ave}} = \cov{\btheta},
\end{equation}
i.e., that the covariance of the parameters reconstructed from $\br$ and $\br_\mathrm{ave}$ are the same, which implies that reconstruction is equally (un)successful with and without application of BDA.

To re-iterate, BDA does not affect our ability to reconstruct an image from the visibility data as long as
\begin{enumerate}
\item the relation between the source and noise signals and the visibilities can be described by a linear transformation;
\item the image reconstruction process can be described by a linear transformation $\bM_\mathrm{im}$ whose rows have a structure commensurate with BDA.
\end{enumerate}
It can be easily demonstrated that the latter condition holds for ML image reconstruction using weighted least squares optimisation as discussed above and for formation of dirty images by means of a discrete Fourier transform. The derivation presented in this section holds for a perfectly calibrated array. Calibration usually requires a non-linear model. As a result, the reasoning in this section does not hold. Fortunately, most radio astronomical observations are low-SNR observations making the information loss due to BDA negligible. Also, many radio astronomical systems are very stable requiring only small corrections, which minimises the impact of the non-linearities.

\subsection{Decorrelation}
\label{ssec:decorrelation}

Decorrelation, or amplitude loss, due to time averaging over total integration time $T$ on the baseline with mid-point $(u_0, v_0, w_0)$ for a source located at direction cosine $(l, m, n)$ can be described by the amplitude reduction factor given by Eq.~(18-31) in \citep{Bridle1999}
\begin{eqnarray}
\lefteqn{R_T ~=} \nonumber\\
& = & \mathrm{sinc} \left \{\pi T  \left ( \left. \frac{\partial u}{\partial t} \right |_{u_0} l + \left . \frac{\partial v}{\partial t} \right |_{v_0} m  + \left . \frac{\partial w}{\partial t} \right |_{w_0} (n-1) \right ) \right \} \nonumber\\
& \approx & 1 - \frac{\pi^2 T^2}{6} \left ( \left. \frac{\partial u}{\partial t} \right |_{u_0} l + \left. \frac{\partial v}{\partial t} \right |_{v_0} m \right )^2,
\label{eq:ampl_reduction}
\end{eqnarray}
where the approximation assumes that $\nu T << 1$, and that the direction cosine offset angles are small,  i.e., that the decorrelation is small.

If $L_x$, $L_y$ and $L_z$ denote the baseline components along the principal axes of the ITRF coordinate system, \cite{Bridle1999} show that
\begin{eqnarray}
\frac{\partial u}{\partial t} & = & \frac{1}{\lambda} \left ( L_x \cos H - L_y \sin H \right ) \omega_E\nonumber\\
\frac{\partial v}{\partial t} & = & \frac{1}{\lambda} \left ( L_x \sin\delta \sin H + L_y \sin \delta \cos H \right) \omega_E,
\end{eqnarray}
where $\delta$ and $H$ denote the declination and hour angle of the source and $\omega_E = 7.2925\cdot 10^{-5}$ rad/s is the angular velocity of the Earth.

In the simulations presented in Sec.~\ref{sec:sims}, we will assess the impact of BDA on imaging observations. Natural weighting provides the best sensitivity, in particular for centrally condensed array configurations, while uniform weighting provides a better angular resolution. We will therefore present results for both natural weighting and uniform weighting. For natural weighting, the expected amplitude reduction in the image can be determined by taking the average amplitude reduction factor for all baselines, where the amplitude reduction factor for each baseline is calculated by Eq.~\eqref{eq:ampl_reduction}. For uniform weighting, we need to take into account an appropriate weighting factor per baseline. As this weighting factor depends on the density of $(u,v,w)$-samples, this requires calculation of $\partial w / \partial t$ as well. It is straightforward to derive this from Eq.~(18-32) in \citep{Bridle1999}, resulting in
\begin{equation}
\frac{\partial w}{\partial t} = - \frac{1}{\lambda} \left ( L_x \cos \delta \sin H + L_y \cos \delta \cos H \right ) \omega_E.
\end{equation}

The weighting factor for uniform weighting can now be described by
\begin{equation}
W_\mathrm{uni} = \frac{1}{N} \sqrt{ \left ( \frac{\partial u}{\partial t} \right )^2 + \left ( \frac{\partial v}{\partial t} \right )^2 +  w_\mathrm{factor} \left ( \frac{\partial w}{\partial t} \right )^2}
\end{equation}
where $N$ is the number of baseline data values within a given $(u, v, w)$ distance around the location of the baseline under consideration. In his Appendix E,  \cite{Briggs1995PhD} shows that the maximum $(u,v,w)$ distance can be specified by $\pi^{-0.5}\Delta$ where $\Delta$ would be the corresponding gridded $(u, v)$-plane increment. For a source 2$^\circ$ from the field centre this maximum distance works out to be about 8 wavelengths. Test simulation runs showed that only 4\% of the simulated visibilities had N greater than 1, and just setting $N=1$ gave minimal differences in the attenuation factors derived in the next section, but sped up the computations enormously. 

At small distances from the field centre the ${\partial w}/{\partial t}$ term in the above equation basically contributes nothing but at large distances from the field centre begins to provide a contribution due to the rotation matrices necessary to shift the phase centre to the new location. Empirical tests (see Sec.~\ref{ssec:wf_effects}) showed that giving the $w_\mathrm{factor}$ term a value of about 0.15 gives a good fit to the data. 

BDA can also be applied over frequency. \cite{Bridle1999} show that the effect of frequency smearing can be described as a convolution in the image domain with a distortion function. This distortion function causes a radial broadening of the point source response. Although the integrated flux density of the source is preserved, this broadening causes a reduction of the amplitude proportional to $1 / \left ( \Delta \nu \sqrt{l_0^2 + m_0^2} \right )$. While it would be interesting to include frequency averaging in our simulations, simulated SKA observations at just a single frequency already required significant computing resources. So we will only investigate the effects of time averaging in the simulations described in the next section and therefore not discuss the effects of BDA over frequency in more detail here.

\section{BDA and Calibration}
\label{sec:CEC}

A practical issue with visibility data obtained using BDA is that visibility samples may have different integration times and, as a result, that the time series for different baselines have different lengths. Calibration routines will thus have to carry out appropriate interpolations to account for the different sampling times and appropriate weighting to account for the different integration times. Not all calibration routines may be able to do this. Another possible calibration issue is that the time scales of certain instrumental and environmental variations are such that these variations are well sampled by the longest baselines (with the shortest integration times) while these time scales may be comparable with the longest integration time on the shortest baselines. Both issues can be handled by the Compress-Expand-Compress (CEC) method proposed by \cite{Salvini2017URSI}. As calibration is a major concern for modern radio interferometers, we summarize this method and recapitulate the main results from \cite{Salvini2017URSI} below.

The CEC method consists of three steps:
\begin{description}
\item{\it Compression:} After crosscorrelation and, if desired, flagging, BDA is applied following \eqref{eq:rave_from_r}. This ensures that the amount of data that needs to be transferred to the compute cluster performing calibration (and likely imaging) is reduced.
\item{\it Expansion:} The averaged visibilities are then used to fill the time series of the corresponding visibilities at the full data rate as described by \eqref{eq:rave_to_r}. Calibration can then be carried out at the full time resolution.
\item{\it Compression:} After calibration, the data are compressed again using the desired BDA scheme before imaging. This compression step reduces the amount of data that needs to be gridded. As the short-term variations have been corrected by calibration, the compression at this stage could be even higher than during the first compression stage.
\end{description}

As calibration and imaging may be done on the same computing platform, the benefits of the second compression stage may not be obvious. However, as the first stage of calibration is necessarily either direction-independent calibration or calibration in a very limited number of directions due to the low SNR of the sources observed, we can safely assume that the compute requirements for this initial calibration step are far less than the compute requirements for gridding. The second compression step may thus lead to a significant reduction of computing resources needed for gridding.

A possible concern for the first compression stage is that the temporal variation of the visibility values on, in particular, the short baselines  within the averaging interval may not only be caused by variations in the geometrical delay, but also by temporal variations in some calibration parameters. \cite{Salvini2017URSI} showed that this risk can be mitigated by using not only mid-point averaging (the zeroth order moment) for BDA, but also the first, and possibly even higher, order moments of the visibility data over the averaging interval at the expense of lower data volume reduction during the first compression stage. \cite{Salvini2017URSI} demonstrated with a simulation with extreme instrumental gain variations that the use of first (and second) order moments allows significant compression while retaining the ability to reconstruct 1-$\mu$Jy sources in the vicinity of a 1-Jy source, indicating that over 60 dB dynamic range remains feasible while using BDA in the presence of fast gain variations.

Based on these findings, we decided to focus the simulations presented in Sec.~\ref{sec:sims} on validating the conclusion drawn based on the theoretical analysis presented in Sec.~\ref{sec:theory} that BDA has no detrimental effect on imaging performance besides a predictable level of decorrelation. This allowed us to set up the simulations in such a way that this particular aspect could be isolated from other effects that need to be dealt with when implementing a data processing pipeline for an actual system.

\section{Simulation of Time Averaging Effects}
\label{sec:sims}

\subsection{Description of simulations}

In addition to the theoretical work presented earlier we made simulated observations with the SKA1-mid telescope configuration to check if BDA had direct effect on decorrelation of "actual" observations. We used the SKA1-mid antenna positions as provided by \citet{Heystek2015} plus the locations of the 64 MeerKAT \citep{Jonas2009ProcIEEE} antennas for a total of 197 antennas. With the given antenna positions, the actual physical baseline lengths ranged from 157 km down to 29 m.

We made single channel observations covering the declination range of 10$^\circ$ to -70$^\circ$ in 20$^\circ$ steps for a total of five different declinations. At each declination we observed seven short cut observations of 7 minutes each with an initial correlator dump time of 0.14 seconds on all of the 19306 baselines. The short cut observations were made at hour angles -4.5, -3, -1.5, 0, 1, 2 and 4 hours. Each short observation was simulated and processed separately. The observing frequency was 700 MHz.

The simulation process was as follows:
\begin{itemize}
\item Place a 1-Jy point source at an offset from the field centre of 2$^\circ$ along either direction cosine $m$ (roughly declination) or $l$ (roughly right ascension). Note that an offset of 2$^\circ$ at 700 MHz corresponds to an equivalent offset of 1$^\circ$ at 1400 MHz. We expect these distances to correspond roughly to the 10 percent power response of the dish primary beam and thus to be about the maximum distance from the field centre at which one might expect to do "significant" science.
\item Generate noise-free visibilities with the above mentioned antenna locations and source position using the CASA simulation tool.
\item As the baselines become shorter, average the sampled visibilities according to various criteria (see below) and replace the original "raw" visibilities by their "average" at the same $(u, v)$ location. This allows us to minimize any errors due to different $(u, v)$ coverage. Also, the CASA-based imager that we used, assumes that the WEIGHTS column of a Measurement Set is an indicator of system temperature changes and is not an indicator for how much weight should be assigned to a $(u, v)$ point at the gridding stage of imaging. Obviously, in an actual SKA system, the the $(u, v)$ data would need to be appropriately weighted to represent the $(u, v)$ coverage each averaged point represents.
\item Make images from the averaged visibilities using the extremes of natural and uniform weighting  \citep{Briggs1995PhD}. Natural weighting essentially weights the $(u, v)$ grid proportional to the signal collected in a given area whereas uniform weighting essentially normalizes the signal in each area. In the current SKA1 design, there are many more visibilities at short spacings than long ones, so natural weighting emphasizes short spacings.
\item Determine the difference between the maximum value found in the images and the expected 1-Jy signal as both a function of hour angle and declination. We could keep the size of the images small ($1024 \times 1024$ pixels) by phase-rotating the reference position of the image to the position of the offset target source.
\end{itemize}

Note that in the following plots signals that are minimized at hour angle zero are due to the test source offset in direction cosine $l$ while signals that are maximized at hour angle zero are due to the test source offset in direction cosine $m$. The 6-hour phase offset in time is not surprising as the sources are offset in the sky by exactly 90 degrees.

\subsection{Effects due to correlator integration}

\begin{figure*}
\centering
\begin{minipage}[b]{.49\textwidth}
\includegraphics[width=0.9\columnwidth]{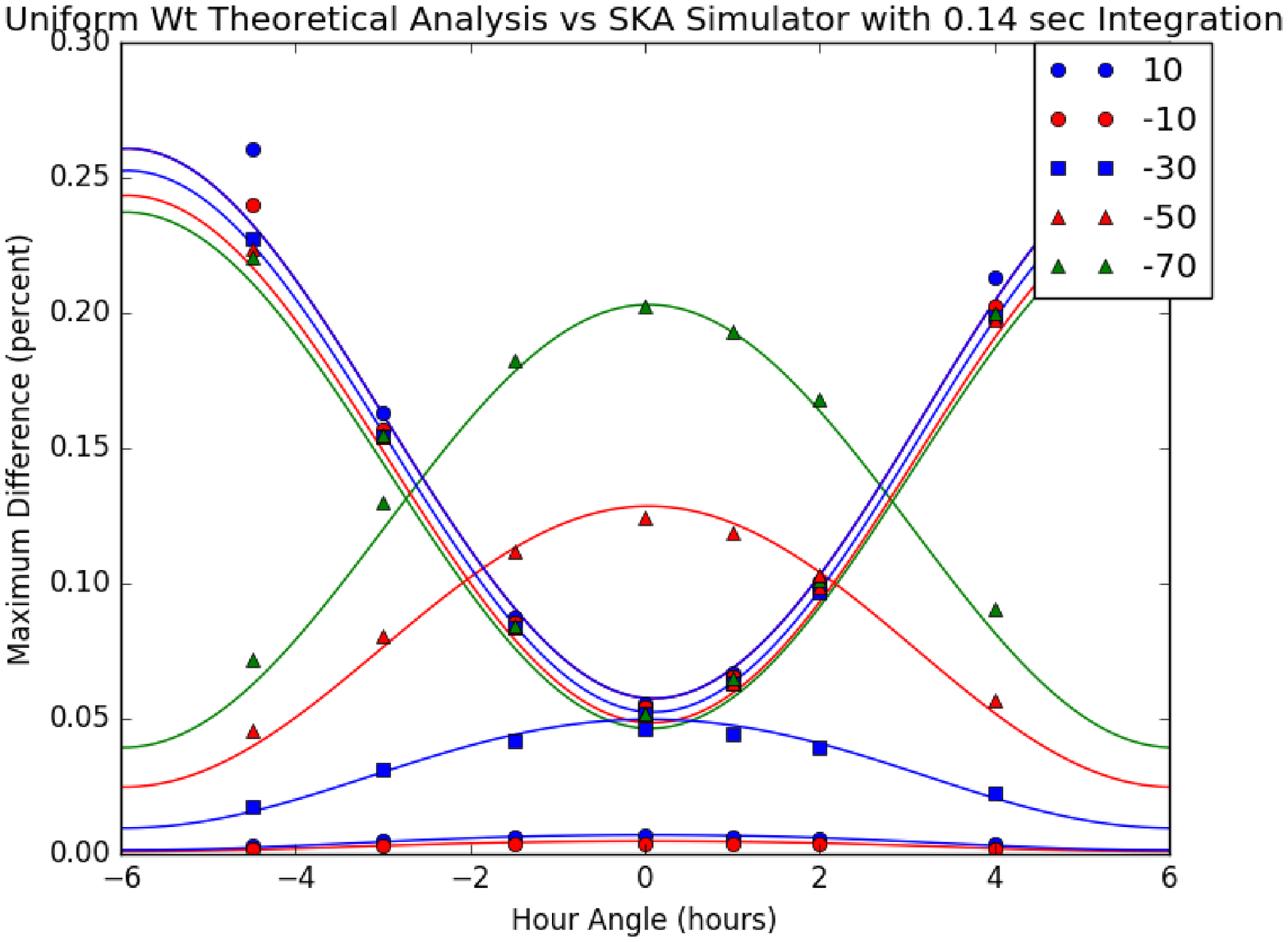}
%\caption{Caption}\label{label-a}
%\end{minipage}\qquad
\end{minipage}\hfill
\begin{minipage}[b]{.49\textwidth}
\includegraphics[width=0.9\columnwidth]{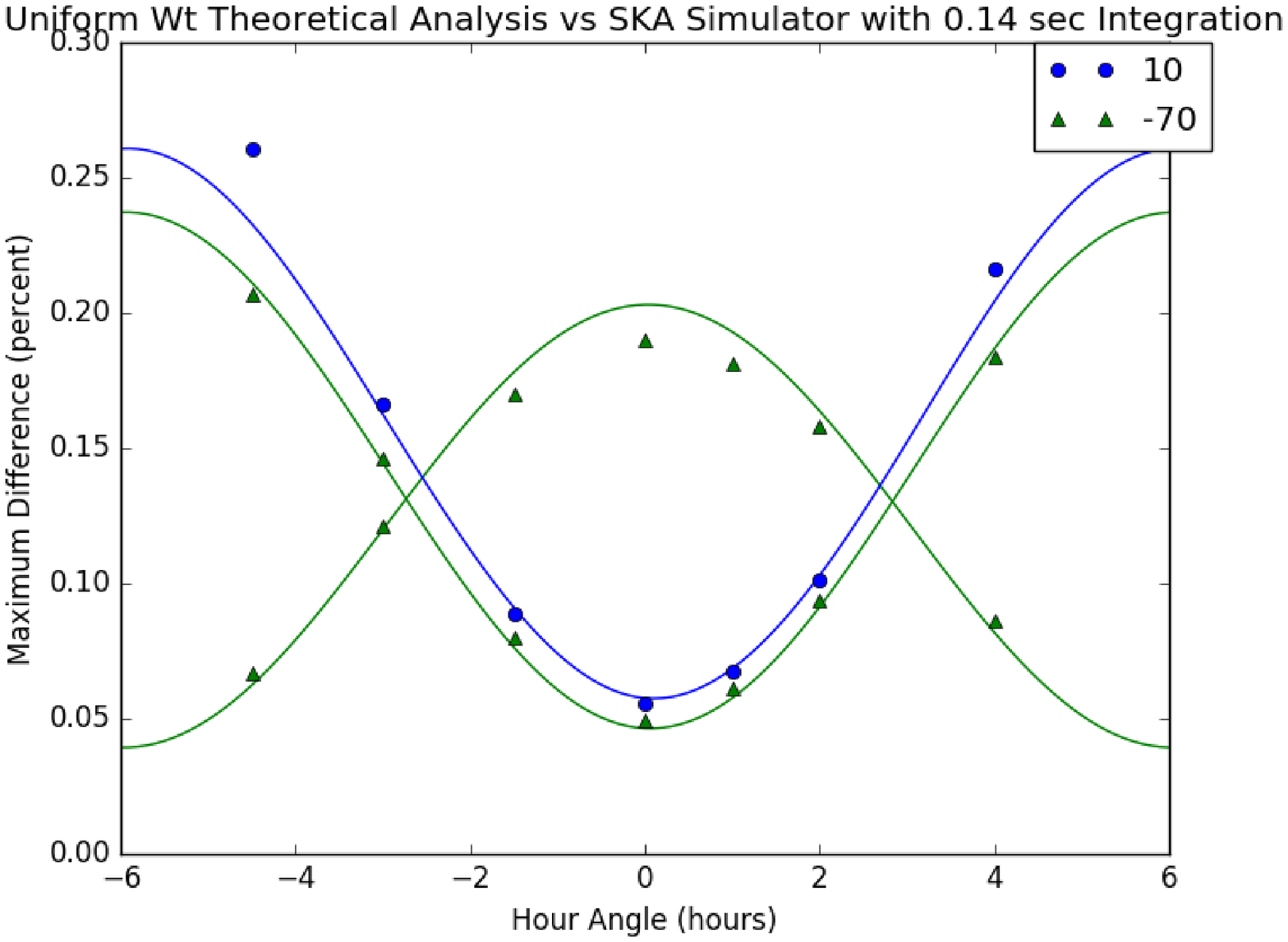}
%\caption{Caption}\label{label-b}
\end{minipage}
\caption{These plots show decorrelation effects on 1-Jy point sources situated 2$^\circ$ from the phase reference position at 700 MHz due to a correlator dump time of 0.14 s. In these and succeeding plots, the residual errors peak near hour angle 0 for a source placed at an offset of 2$^\circ$ in direction cosine $m$ and the errors are minimized near hour angle 0 for a source located at an offset of 2$^\circ$ in direction cosine $l$. The actually measured data points are shown as separate symbols, while the solid lines represent the theoretically predicted attenuation based on the prescription in Sec.~\ref{ssec:decorrelation}. We obtained these results by dividing each original $(u, v)$ step into 20 subintervals and then averaging the calculated amplitudes and phases over the 20 subintervals. While the effects are very small for natural weighting (less than about 0.04 percent) they increase to a maximum of about 0.25 percent for uniform weighting. The left plot shows the maximum residuals in images made with uniform weighting as a function of hour angle and declination. The right plot shows the equivalent results found when we generated actual $(u, v)$ positions with a correlator sample time of 0.007 s and then averaged the data together in blocks of 0.14 s. To obtain the right plot we only calculated the results expected for the most extreme variations as the simulations took about 15 hours of computer time for each data point. For this reason, we did not bother to do simulations at 0.007s resolution for the source offset in direction cosine $m$ at 10$^\circ$ declination. The errors are already very small as shown in the left plot. Decorrelation is expressed as the fractional reduction of power (measured as difference between the original power and observed power) in the original 1-Jy signal. The two approaches give good agreement and validate our approach of linearly interpolating between original $(u, v)$ points to simulate the correlator effects. 
\label{fig:correlator}}
\end{figure*}

The current SKA1 design calls for the correlator to collect data for 0.14 seconds before dumping the result. We find that even this short integration time could have a significant effect on imaging.    
We determined the magnitude of this effect by taking the $(u, v, w)$ locations as calculated at 0.14-s intervals, linearly interpolating them into 20 subintervals, calculating the expected amplitudes and phases at the subintervals and then averaging them together. The results are shown in Fig.~\ref{fig:correlator}. In addition to the results of the simulated observations, the plots also show (solid lines) a prediction of the expected decorrelation for uniform  weighting as a function of hour angle and declination based on the prescription provided in Sec.~\ref{ssec:decorrelation}.

We checked that these results made sense by actually making some simulations where we lowered the correlator dump time to 1/20th of 0.14 s or 0.007 s and then averaged up the signals in 0.14-s blocks on all baselines. We got very similar results. While a residual of 0.25\% sounds small, 0.25\% of a 10-mJy source equates to 25 $\mu$Jy, a signal expected to be well above the SKA signal sensitivity limit.

%\begin{figure}
%\centering
%\includegraphics[width=0.48\columnwidth]{Figures/baselines_alg1.eps}
%\includegraphics[width=0.48\columnwidth]{Figures/baselines_alg2.eps}
%\caption{These plots show the distribution of physical baselines that fall into
%different averaging goups for the first and second algorithms that we discuss. 
%The first algorithm (left figure) tends to distribute the averaging over the 
%entire range of physical baselines. The second averaging algorithm ignores
%the longer baselines with an increased emphasis on the short baselines. The
%initial bar indicates the number of baselines between zero km and the start (in
%km) of the next bin, etc. The final bin marks the limit (in km) beyond which 
%no averaging is done.
%\label{fig:baselines}}
%\end{figure}

\subsection{A simple averaging scheme}

Since our physical baseline length ranges from 157 km down to 29 m, a simple way to average is to double the integration time for every factor 2 decrease in baseline length. This leads to the following baselines zone limits and associated baseline averaging:
\begin{enumerate}
\item [physical baseline zones (km)]~\\ \{ 80, 40, 20, 10, 5, 2.5, 1.25, 0.625, 0.3125 \}
\item [baseline averaging (multiples of correlator dump time)]~\\ \{ 1, 2, 4, 8, 16, 32, 64, 128, 256, 512 \},
\end{enumerate}
which means that baselines longer than 80 km are not averaged, baselines with length between 40 and 80 km have two points averaged, baselines with length between 40 and 20 km have four points averaged, and so on. 

\begin{figure*}
\centering
\begin{minipage}[b]{.49\textwidth}
\includegraphics[width=0.9\columnwidth]{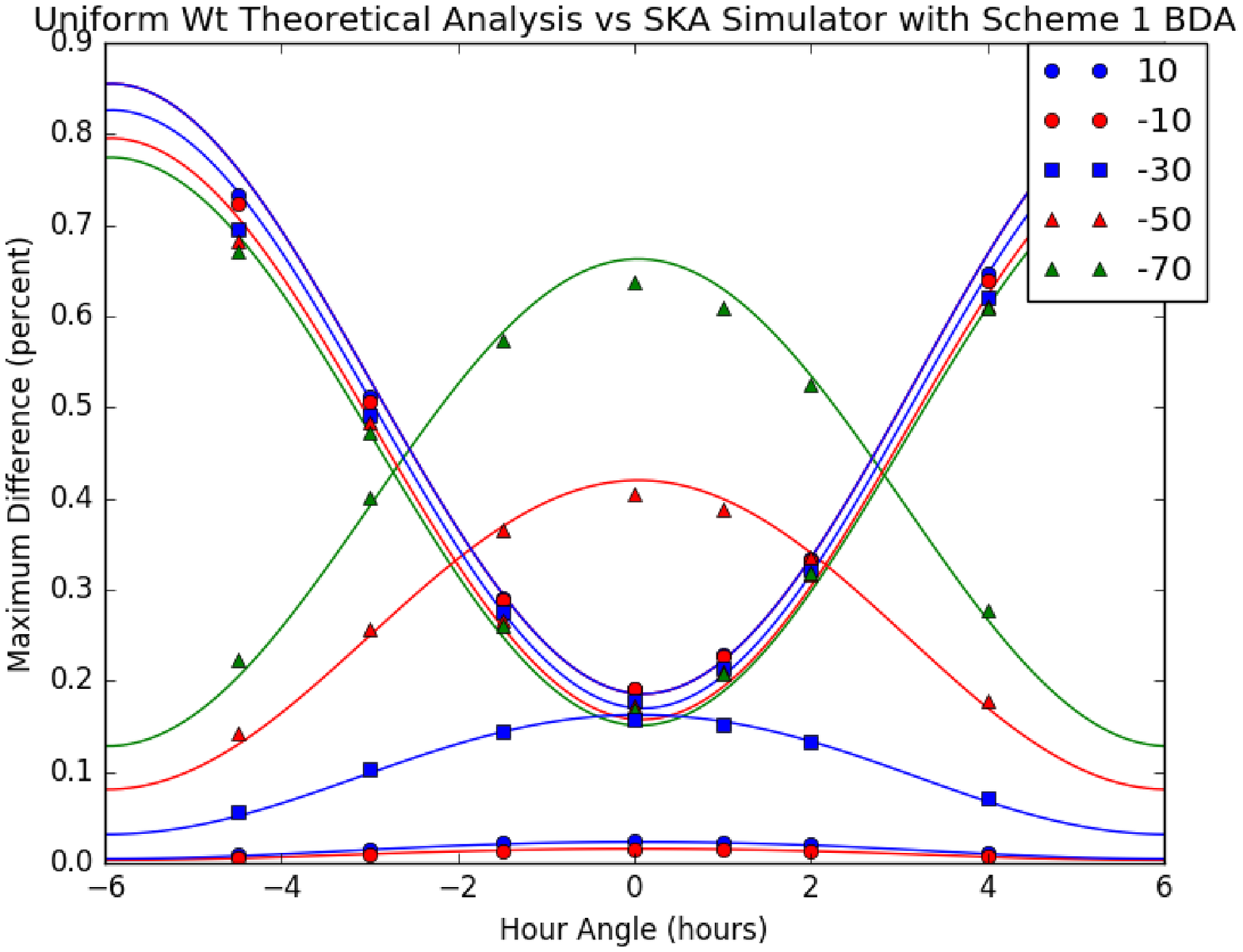}
%\caption{Caption}\label{label-a}
%\end{minipage}\qquad
\end{minipage}\hfill
\begin{minipage}[b]{.49\textwidth}
\includegraphics[width=0.9\columnwidth]{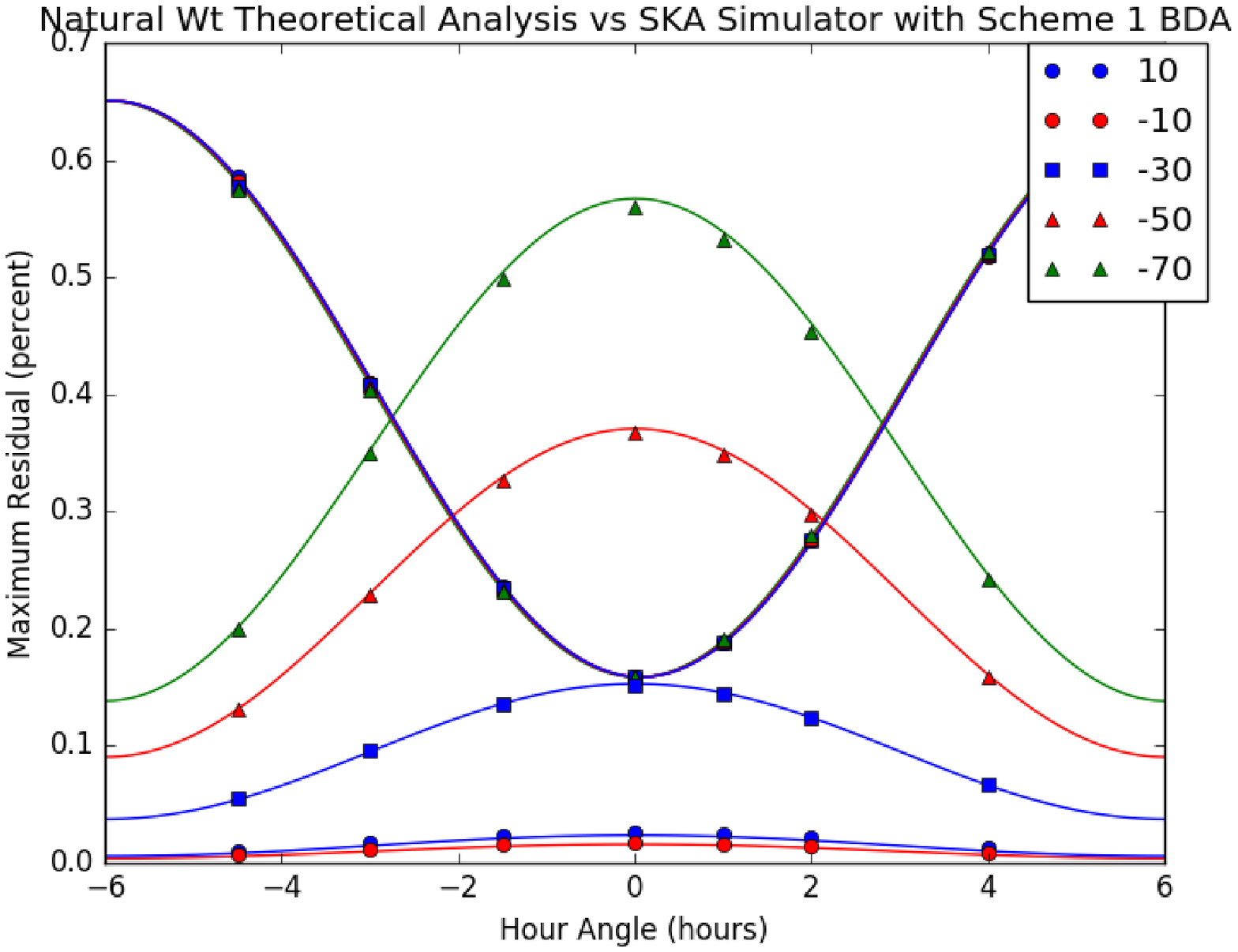}
%\caption{Caption}\label{label-b}
\end{minipage}
\caption{These plots show results for a 1-Jy point source placed at offsets of 2$^\circ$ in direction cosine $m$ (residual errors peak at hour angle 0) and at 2$^\circ$ in direction cosine $l$ (residual errors are minimized at hour angle 0). In the left plot we show the maximum residuals in images made with uniform weighting as a function of hour angle and declination. The right plot shows the equivalent results from images made with natural weighting. Note that the behaviour is very similar. Decorrelation is expressed as a percentage of the original test source signal of 1 Jy.
\label{fig:willis_algorithm}}
\end{figure*}

This scheme gave an 87.5\% reduction in data. However we found that the following scheme:
\begin{enumerate}
\item [physical baseline zones (km)]~\\ \{ 80, 40, 30, 20, 15, 10, 7.5, 5, 3.75, 2.5, 1.875, 1.25, 0.9375, 0.625, 0.5625, 0.375, 0.28125 \}
\item [baseline averaging (multiples of correlator dump time)]~\\ \{ 1, 2, 3, 4, 6, 8, 12, 16, 24, 32, 48, 64, 96, 128, 192, 256, 384, 512 \},
\end{enumerate}
where we inserted some intermediate binning intervals, gave almost as large a reduction in visibility data volume, 86.8\%, but decreased the residual errors. For example, the maximum residual error decreased from 0.73\% to 0.64\% at hour angle zero for the source offset 2$^\circ$ in direction cosine $m$ at declination -70$^\circ$ (see Fig.~\ref{fig:willis_algorithm}). We will refer to this BDA scheme as Scheme 1.

In this analysis, we used actual physical baseline lengths rather than projected $(u, v)$ lengths. The maximum residuals as a function of hour angle and declination for a test source offset by 2 degrees in direction cosines $l$ and $m$ are shown in Fig.~\ref{fig:willis_algorithm}. Note that the observed values and theoretical predictions (solid curves) shown in the figures implicitly include the effects of the 0.14-s correlator integration time.

%\begin{table}
%\centering
%\begin{tabular}{c|c|c}
%\hline
%\hline
%baseline & compression & number of baselines\\
%\hline
%$<$ 1.25 km & 128 & 7196\\
%$<$ 2.5 km & 64 & 2112\\
%$<$ 5 km & 32 & 2561 \\
%$<$ 10 km & 16 & 1544 \\
%$<$ 20 km & 8 & 1673 \\
%$<$ 40 km & 4 & 1891 \\
%$<$ 80 km & 2 & 1737\\
%$<$ 160 km & 0 & 612\\
%\hline
%\hline
%\end{tabular}
%\caption{First Algorithm: Compression values used for different baseline ranges. \label{tab:algorithm_one}}
%\end{table}

%\begin{figure}
%\centering
%\includegraphics[width=0.9\columnwidth]{Figures/willis_uniform_weighting_BDA_plot.eps}
%\includegraphics[width=0.9\columnwidth]{Figures/willis_natural_weighting_BDA_plot.eps}
%\caption{These plots show results from the first averaging scheme for  
%a 1 Jy point source placed at offsets of 2$^\circ$ in direction cosine $m$ (resudual errors peak at 0 hour hour angle) and at 2$^\circ$ in direction cosine $l$ (resuduals errors are minimized at hour angle 0 degrees) .
%In the left plot we show the maximum residuals in images made with uniform 
%weighting s a function of hour 
%angle and declination The right plot shows the equivalent results from
%images made with natural weighting. Note that the behaviour is very similar.
%Units are given as a percentage of the original test source signal of 1 Jy.
%\label{fig:willis_algorithm}}
%\end{figure}

Despite the fact that we have been averaging data from a large number of baselines with a wide variety of orientations, the behaviour of the residuals shown in Fig.~\ref{fig:correlator} and Fig. \ref{fig:willis_algorithm} is astonishingly consistent with the simple analysis of time-averaging given on pages 206 through 208 of \citep{Thompson2001}. At extreme hour angles, the corrugations of interferometer fringes are parallel to the $l$-direction cosine but the Earth's rotation vector direction is perpendicular to the $l$-direction and $l$-fringe corrugations. So a source located along the $l$-direction cosine will be moving through fringes most rapidly and be most susceptible to averaging effects, whereas the Earth's rotation causes a source along the $m$-direction to be moving approximately parallel to fringe corrugations so averaging has little effect. The inverse situation occurs at transit. \cite{Thompson2001} also show (their equation 6.81) that residual effects in $m$ should be subject to a $\sin^2(\mathrm{dec})$ effect, and indeed the increase in $m$ residuals at transit follows quite closely $\sin^2 (\mathrm{dec})$ behaviour.

Although \cite{Thompson2001} did not derive any declination dependent effects for a source offset in direction cosine $l$ our results do show a small effect. We believe that this is due to the fact that, in order to make small 1024 x 1024 pixel images. we have to do a rotation of the phase centre to the position of the offset source. This requires a sequence of rotation matrices which will cause the $w$-coordinate to start contributing to the $u$ and $v$ velocity terms. 

\begin{figure}
\centering
\includegraphics[width=0.9\columnwidth]{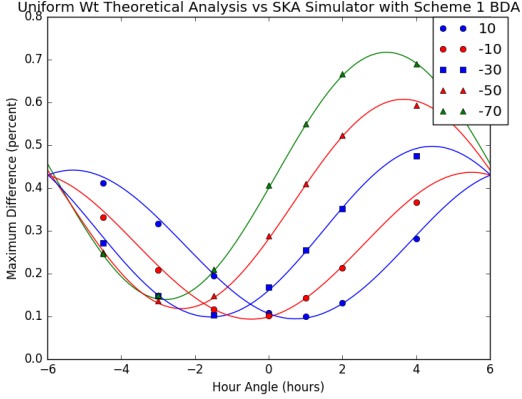}
\caption{This figure shows the predicted and observed decorrelation as a function of hour angle when the test source is placed at an ofset of 1.414 degrees in both $l$ and $m$ direction coordinates. \label{fig:diagonal}}
\end{figure}

The plots shown so far show a symmetry with respect to hour angle. However be reminded that this is simply due to the special positions at which we have placed the test source. If we put the test source at an offset of 1.414 degrees in both $l$ and $m$ direction coordinates, we get an asymmetric effect as shown in Fig.~\ref{fig:diagonal}. Essentially every source in the field will follow a different pattern.

\begin{figure}
\centering
\includegraphics[width=0.9\columnwidth]{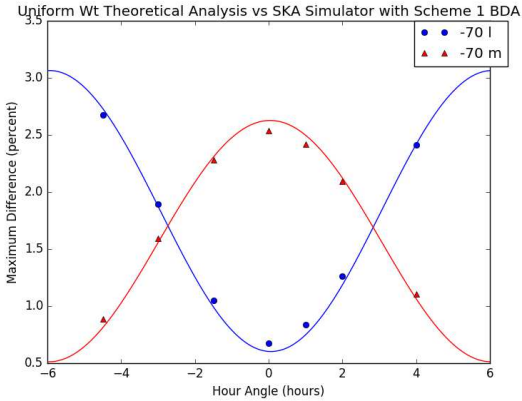}
\caption{This figure shows the predicted and observed decorrelation as a function of hour angle for Scheme 1 BDA but with double integration time on all baselines. We get rid of 93\% of the data (from 87\% in the original scheme) while the maximum residual error increases to about 2.6\% from about 0.65\% previously.\label{fig:double_bda}}
\end{figure}

By using Scheme 1 BDA we managed to shrink the amount of data by 87\%. If we double the averaging period on all baselines (so even on the longest baselines we now average two points together) we can even reduce the amount of visibility data by 93\% at the expense of increasing the residual errors to 2.6\% (see Fig.~\ref{fig:double_bda}). This error may still be acceptable for SKA data processing at large distances from the field centre.

\begin{figure}
\centering
\includegraphics[width=0.9\columnwidth]{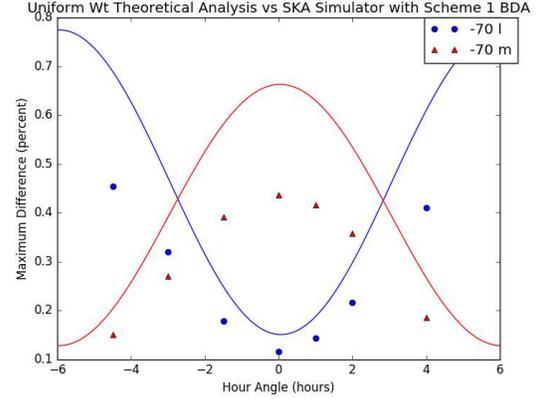}
\caption{This figure shows the predicted and observed decorrelation as a function of hour angle when a source is inverted into the $(u, v)$-plane and BDA is done but the 0.14-s nominal correlator dump time is ignored. \label{fig:clean_comp}}
\end{figure}

Another interesting effect is that even the nominal correlator dump time must be included in predictions related to BDA. In Fig.~\ref{fig:clean_comp} we show the effect of having inverted a 1-Jy source into the $(u, v)$-plane and applying our BDA scheme but without adjusting for the internal 0.14 s integration inside the correlator. It is clear that the effects of BDA are underestimated.

\subsection{Wide Field Effects}
\label{ssec:wf_effects}

\begin{figure}
\centering
\includegraphics[width=0.9\columnwidth]{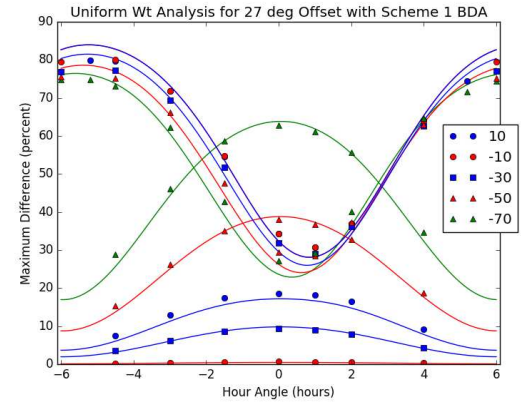}
\caption{This figure shows the predicted and observed decorrelation as a function of hour angle and declination when uniform weighting is used to image a source situated 27.2$^\circ$ from the field centre. \label{fig:wide_field_effects}}
\end{figure}

Our previous simulations were restricted to sources within a few degrees of the field centre. However, as radio interferometer sensitivity increases, there is an increasing possibility that strong background sources will appear above the background noise out to great distances from the central field of view. Out of curiosity we placed a source at 27.2$^\circ$ from the field centre (so roughly 13 primary beams at 700 MHz) and applied the BDA averaging scheme described in this section to the visibilities. At this distance we have to use the full $\mathrm{sinc}$ decorrelation function given in the first part of Eq.~\eqref{eq:ampl_reduction} when calculating the expected decorrelation. The results are shown in Fig.~\ref{fig:wide_field_effects}. The actual simulated observations show excellent agreement with the theoretical predictions and suggest that the techniques discussed here could be adapted to remove the effects of such sources from actual observations.

\subsection{Second averaging scheme}

\begin{figure*}
\centering
\begin{minipage}[b]{.49\textwidth}
\includegraphics[width=0.9\columnwidth]{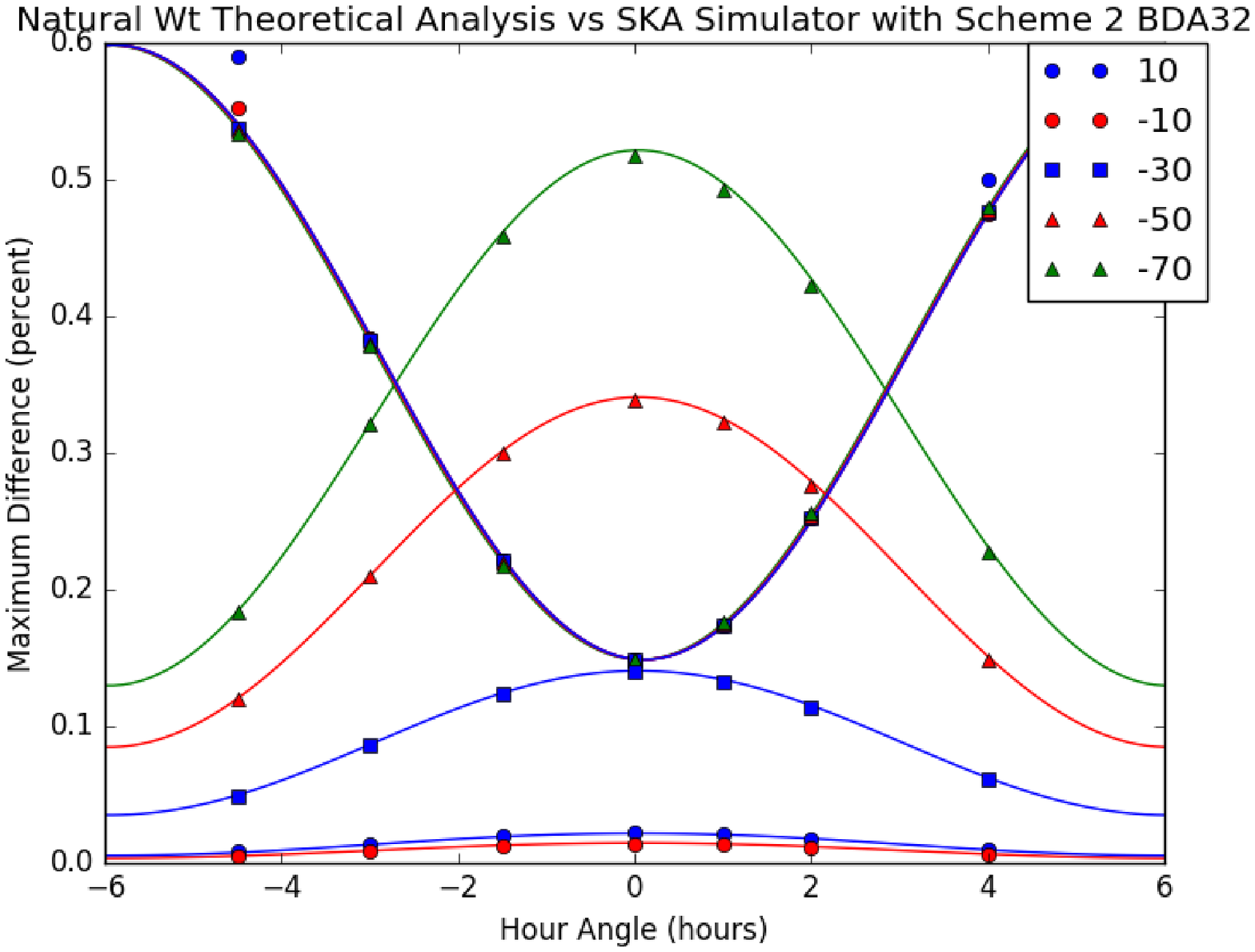}
%\caption{Caption}\label{label-a}
%\end{minipage}\qquad
\end{minipage}\hfill
\begin{minipage}[b]{.49\textwidth}
\includegraphics[width=0.9\columnwidth]{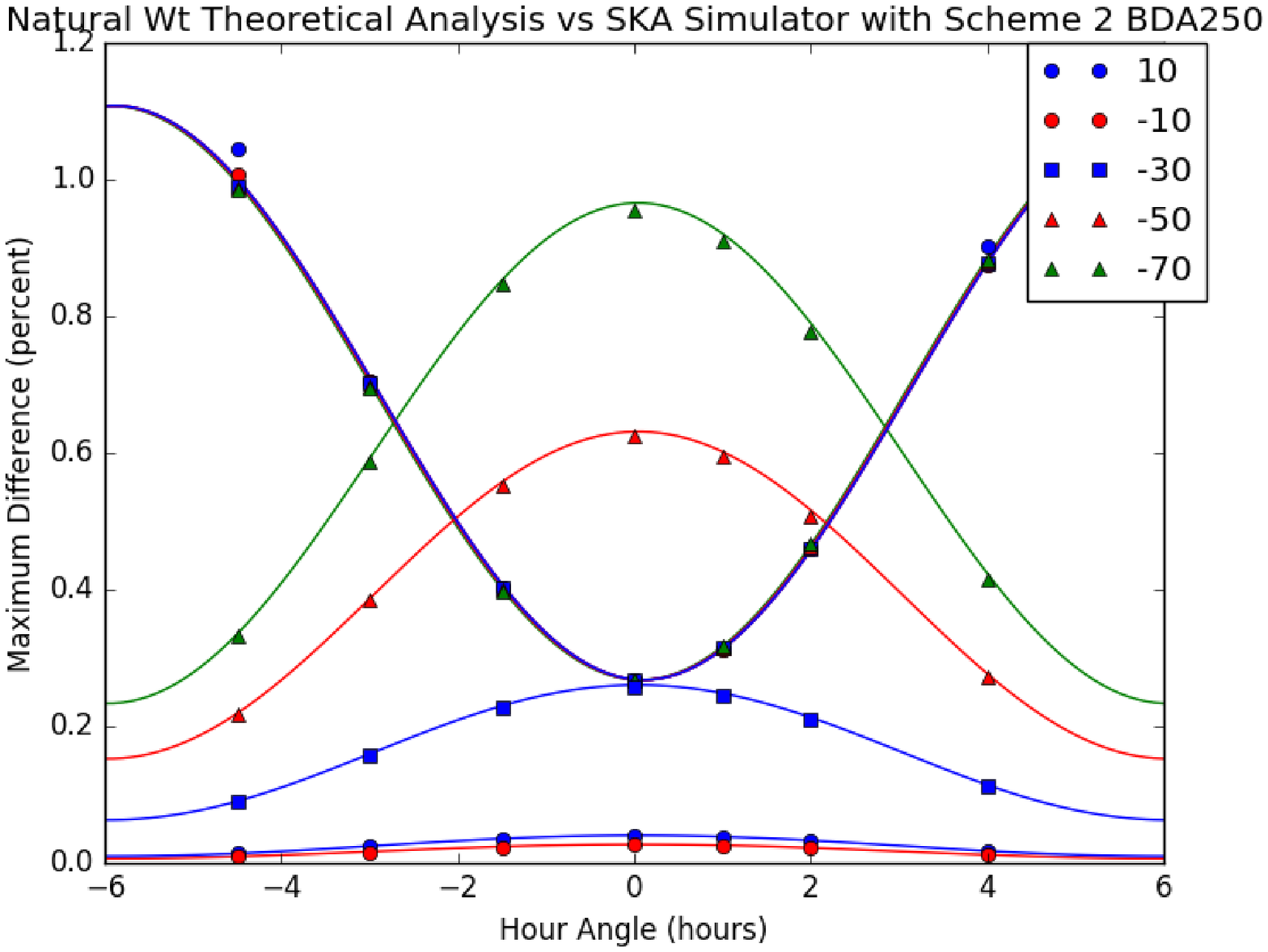}
%\caption{Caption}\label{label-b}
\end{minipage}
\caption{These plots show results for BDA Scheme 2. The plots show the variation as a function of hour angle and declination for a unit point source placed at an offset of 2$^\circ$ in direction cosine $l$ and $m$ from the field centre. The left plot shows the maximum residual in the images made with natural weighting and a maximum averaging value of 32. The right plot shows the corresponding residual behaviour for a maximum averaging value of 250. 
\label{fig:salvini_algorithm_natural}}
\end{figure*}

A potential disadvantage of BDA Scheme 1 proposed above is that doubling the integration time every time the baseline length halves may cause sharp changes in the level of decorrelation at specific spatial scales. To smooth the variation of decorrelation with spatial scale even further, we developed a second scheme. This BDA Scheme 2 works as follows: define a maximum allowable number of points to average. Then the averaging value for a given baseline becomes the minimum of the maximum allowable value and the integer value of the maximum baseline length divided by the current baseline length. So we average baselines up to a maximum allowable value and all shorter baselines are averaged to this maximum value. This scheme was used by \citet{Salvini2017URSI} to test the CEC method proposed in that paper and summarized in Sec.~\ref{sec:CEC}. We varied the maximum allowable value from 32 to 500. The corresponding amount of data compression ranged from 87.8\% to 89.1\%, so there was rather little gain in allowing for very long integration times on the shortest baselines. In Fig.~\ref{fig:salvini_algorithm_natural} we show the results for natural weighting obtained with this scheme, which confirm the earlier conclusions drawn based on our simulations with BDA Scheme 1.

%\begin{table}
%\centering
%\begin{tabular}{c|c|c}
%\hline
%\hline
%baseline & compression & number of baselines\\
%\hline
%%$<$ 234.38 m & 768 & 1673\\
%%$<$ 468.75 m & 384 & 2158\\
%$<$ 468.75 m & 384 & 3831\\
%$<$ 937.5 m & 192 & 2427\\
%$<$ 1.875 km & 96 & 2075 \\
%$<$ 3.75 km & 48 & 2440\\
%$<$ 7.5 km & 24 & 2073 \\
%$<$ 15 km & 12 & 1530\\
%$<$ 30 km & 6 & 1621\\
%$<$ 160 km & 0 & 3309\\
%\hline
%\hline
%\end{tabular}
%\caption{Second Algorithm: Compression values used for different baseline ranges. \label{tab:algorithm_two}}
%\end{table}

%\begin{figure*}
%\centering
%\begin{minipage}[b]{.49\textwidth}
%\includegraphics[width=0.9\columnwidth]{Figures/salvini_uniform_weighting_BDA32.eps}
%\end{minipage}\hfill
%\begin{minipage}[b]{.49\textwidth}
%\includegraphics[width=0.9\columnwidth]{Figures/salvini_uniform_weighting_BDA250.eps}
%\end{minipage}
%\caption{These plots show results for the second averaging scheme. The plots shows the variation as a function of hour angle and declination for a unit point source placed at an offset of 2 degrees in direction cosine $l$ and $m$ from the field centre. The left plot shows the maximum residual in images made with uniform weighting and a maximum averaging value of 32. The right plot shows the corresponding residual behaviour for a maximum averaging value of 250.
%\label{fig:salvini_algorithm_uniform}}
%\end{figure*}

One thing that is very noticeable in these results is that the residuals increase as the maximum compression value increases. The natural weighting residuals for a maximum averaging value of 32 agree well with those found for the first BDA scheme but become worse as the maximum averaging value increases as can be seen by comparing the plots from Figure \ref{fig:salvini_algorithm_natural} with the right hand plot of Figure \ref{fig:willis_algorithm}.
%Also, while our theoretical analysis agrees well with the natural weighting results for this second aveeraging scheme, the observed results for uniform weighting are clearly higher than those predicted from our theoretical model as is shown in Figure \ref{fig:salvini_algorithm_uniform}.

The BDA schemes described above reduce the visibility data volume by at least 87\%. We tested different averaging schemes and found that schemes with a similar level of decorrelation give a similar reduction in data volume. We suggest that BDA can significantly decrease the amount of SKA data that would need to be sent through SKA gridding and imaging pipelines. The residual errors of the suggested schemes are less than one percent at the edge of the primary beam field-of-view. An advantage of using an averaging scheme based on the physical baseline length is that for extensive surveys covering a large region of sky the data are processed in a uniformly consistent way. 

%\begin{table}
%\begin{center}
%    \begin{tabular}{| l | l | }
%    \hline
%    \multicolumn{2}{|c|}{Data decrease for two averaging schmes} \\
%    \hline
%    Scheme 1 & 87 \% \\ \hline
%    Scheme 2 &  88 to 89 \% \\ \hline
%    \end{tabular}
%\end{center}
%\caption{Averaging summary \label{tab:averaging}}
%\end{table}

\section{Conclusions}

In this paper, we presented a theoretical analysis indicating that baseline dependent averaging (BDA) does not impair the imaging capabilities of the instrument. If significant calibration corrections need to be made, some information is lost, but our analysis indicates that this only affects observations with a high instantaneous SNR, i.e., observations of sources whose source temperature is comparable with the system temperature. In such cases, this loss is probably acceptable. 

These theoretical findings were confirmed by simulations, which did not show any indication of other detrimental effects than well-understood decorrelation. Our simulations indicate that by using BDA over time only, the visibility data volume of the SKA1-mid telescope can be reduced by over 87\% while keeping the decorrelation of sources at the edge of the field-of-view below 1\%. Even more aggressive visibility data reduction could be achieved for continuum observations by applying BDA along the frequency dimension as well.

\section{Acknowledgement}

The authors would like to thank Jeroen Stil, Richard Dodson, Daniel Mitchell, Benjamin Mort and Fred Dulwich for the useful suggestions and comments they made while we were working on the material presented in this paper.

\bibliographystyle{mn2e}

\label{lastpage}

\end{document}